\documentclass[%
 reprint,
 superscriptaddress,
groupedaddress,
 amsmath,amssymb,
 aps,
]{revtex4-2}

\usepackage{graphicx}
\usepackage{dcolumn}
\usepackage{bm}

\begin{document}

\preprint{LBNL Tech.\ Report}

\title{Whitepaper submitted to Snowmass21:\\
Advanced accelerator linear collider demonstration facility at intermediate energy }
\author{C. Benedetti}
\affiliation{Lawrence Berkeley National Laboratory, Berkeley, California 94720, USA}
\author{R. Bernstein}
\affiliation{Fermi National Accelerator Laboratory, P.O. Box 500, Batavia, IL 60510, USA}
\author{S. S. Bulanov}
\email[Corresponding author: ]{SBulanov@lbl.gov}
\affiliation{Lawrence Berkeley National Laboratory, Berkeley, California 94720, USA}
\author{E. Esarey}
\affiliation{Lawrence Berkeley National Laboratory, Berkeley, California 94720, USA}
\author{C. G. R. Geddes}
\affiliation{Lawrence Berkeley National Laboratory, Berkeley, California 94720, USA}
\author{S. J. Gessner}
\affiliation{SLAC National Accelerator Laboratory, Menlo Park, California 94025, USA}
\author{A. J. Gonsalves}
\affiliation{Lawrence Berkeley National Laboratory, Berkeley, California 94720, USA}
\author{M. J. Hogan}
\affiliation{SLAC National Accelerator Laboratory, Menlo Park, California 94025, USA}
\author{P. M. Jacobs}
\affiliation{Lawrence Berkeley National Laboratory, Berkeley, California 94720, USA}
\author{C. Jing}
\affiliation{Argonne National Laboratory, Lemont, Illinois 60439, USA}
\author{S. Knapen}
\affiliation{Lawrence Berkeley National Laboratory, Berkeley, California 94720, USA}
\author{I. Low}
\affiliation{Argonne National Laboratory, Lemont, Illinois 60439, USA}
\author{C. Lee}
\affiliation{Theoretical Division, Los Alamos National Laboratory, Los Alamos, NM 87545, USA}
\author{X. Lu}
\affiliation{Argonne National Laboratory, Lemont, Illinois 60439, USA}
\author{P. Meade}
\affiliation{Stony Brook University, Department of Physics and Astronomy, Stony Brook, New York 11794, USA}
\author{P. Muggli}
\affiliation{Max Planck Institute for Physics, 80805 Munich, Germany}
\author{P. Musumeci}
\affiliation{University of California, Los Angeles, California 90095, USA}
\author{B. Nachman}
\affiliation{Lawrence Berkeley National Laboratory, Berkeley, California 94720, USA}
\author{K. Nakamura}
\affiliation{Lawrence Berkeley National Laboratory, Berkeley, California 94720, USA}
\author{T. Nelson}
\affiliation{SLAC National Accelerator Laboratory, Menlo Park, California 94025, USA}
\author{S. Pagan Griso}
\affiliation{Lawrence Berkeley National Laboratory, Berkeley, California 94720, USA}
\author{M. Palmer}
\affiliation{Accelerator Test Facility, Brookhaven National Laboratory, Upton, New York 11973, USA}
\author{E. Prebys}
\affiliation{University of California, Davis, California 90095, USA}
\author{C. B. Schroeder}
\affiliation{Lawrence Berkeley National Laboratory, Berkeley, California 94720, USA}
\author{V. Shiltsev}
\affiliation{Fermi National Accelerator Laboratory, P.O. Box 500, Batavia, IL 60510, USA}
\author{D. Terzani}
\affiliation{Lawrence Berkeley National Laboratory, Berkeley, California 94720, USA}
\author{J. van Tilborg}
\affiliation{Lawrence Berkeley National Laboratory, Berkeley, California 94720, USA}
\author{M. Turner}
\affiliation{Lawrence Berkeley National Laboratory, Berkeley, California 94720, USA}
\author{N. Vafaei-Najafabadi}
\affiliation{Stony Brook University, Department of Physics and Astronomy, Stony Brook, New York 11794, USA}
\author{W.-M. Yao}
\affiliation{Lawrence Berkeley National Laboratory, Berkeley, California 94720, USA}
\author{R. Yoshida}
\affiliation{Argonne National Laboratory, Lemont, Illinois 60439, USA}
\author{L. Visinelli}
\affiliation{Tsung-Dao Lee Institute (TDLI) \& School of Physics and Astronomy, \\ Shanghai Jiao Tong University, Shanghai, China}
\author{C. A. Aidala} 
\affiliation{Department of Physics, University of Michigan, Ann Arbor, Michigan 48109, USA}
\author{A. G. R. Thomas}
\affiliation{Center for Ultrafast Optical Science, University of Michigan, Ann Arbor, Michigan 48109, USA}

\date{\today}

\maketitle

\onecolumngrid
\newpage
\section*{Executive Summary}

It is widely accepted that the next lepton collider beyond a Higgs factory would require center-of-mass energy of the order of up to 15 TeV. Since, given reasonable space and cost restrictions, conventional accelerator technology reaches its limits near this energy, high-gradient advanced acceleration concepts are attractive. Advanced and novel accelerators (ANAs) are leading candidates due to their ability to produce acceleration gradients on the order of 1--100~GV/m, leading to compact acceleration structures. Over the last 10-15 years significant progress has been achieved in accelerating electron beams by ANAs. For example, the demonstration of several-GeV electron beams from laser-powered capillary discharge waveguides, as well as the proof-of-principle coupling of two accelerating structures powered by different laser pulses, has increased interest in ANAs as a viable technology to be considered for a compact, TeV-class, lepton linear collider.

However, intermediate facilities are required to test the technology and demonstrate key subsystems. A 20-100 GeV center-of-mass energy ANA-based lepton collider can be a possible candidate for an intermediate facility. Apart from being a test beam facility for accelerator and detector studies, this collider will provide opportunities to study muon and proton beam acceleration, investigate charged particle interactions with extreme electromagnetic fields (relevant for beam delivery system designs and to study the physics at the interaction point), as well as precision Quantum Chromodynamics and Beyond the Standard Model physics measurements. Possible applications of this collider include the studies of $\gamma\gamma$ and $e$-ion collider designs.  

The advanced accelerator and HEP communities propose the following recommendations to the Snowmass conveners:

\begin{enumerate}
\item The research continue on the science case for the intermediate facility in the framework of the General Accelerator R\&D (GARD) program.

\item A design to be carried out for a  collider demonstration facility at an intermediate energy (20-100~GeV)  to test the technology and demonstrate key subsystem, as well as  provide a facility for physics experiments at intermediate energy.

\end{enumerate}

\newpage
\twocolumngrid

\tableofcontents
\section{\label{sec:intro} Introduction}

The two most important parameters of a linear collider are the maximum beam energy and the luminosity. They are limited by the accelerating gradient for a given accelerator technology and the power cost. For example, the proposed Compact Linear Collider (CLIC) relies on conventional metallic RF structures with a peak accelerating gradient of 100 MV/m, and the size of the linac to reach TeV-class center-of-mass energies is tens of kilometers \cite{CLIC-CDR}. Advanced and novel accelerators (ANAs) \cite{Esarey09,Hooker13} have demonstrated gradients 10-100 GV/m, orders of magnitude larger than conventional RF accelerators. A linear collider based on ANAs offers the possibility of orders of magnitude reduction in the size of the collider linacs, and the associated reductions in cost. In addition, ANAs naturally generate ultrashort bunches (sub-100 fs), significantly reducing the beamsstrahlung during the collision, and, hence, reducing the required beam power to reach a luminosity goal \cite{Schroeder10b}. 

ANAs may be plasma-based or dielectric-based structures driven by short charged particle beams or by intense, short-pulse lasers that resonantly excite large amplitude wakefields with relativistic phase velocity. Charged particle beams interact with the wakefields in the plasma or dielectric structure gaining energy. There has been tremendous progress in the ANA field in the last two decades. Much of the progress on laser-driven plasma accelerators (LPAs) has been the result of laser technology advances and a more complete understanding of the laser-plasma/laser-structure interaction physics. Electron beams have been accelerated up to 8 GeV over 20 cm at Berkeley Lab using a PW laser pulse propagating in a plasma channel \cite{Gonsalves19}. A discussion of the status of ANA research as well as the necessary avenues for achieving progress in this field may be found in the whitepapers ``Linear collider based on laser-plasma accelerators'', ``AWAKE, Plasma Wakefield Acceleration of Electron Bunches for Near and Long Term Particle Physics Applications'', and ``Near Term Applications driven by Advanced Accelerator Concepts'' submitted to Snowmass21 \cite{LPA-collider,near-term-Snowmass21,AWAKE}.

Given the potential of compact, high gradient, advanced accelerator technologies, the DOE Office of High Energy Physics published a report outlining a R$\&$D roadmap toward a collider \cite{roadmap16}. A significant planning effort is happening concurrently in Europe \cite{allegro.roadmap.2019,EPP.roadmap.2022}. As envisioned in these roadmaps, realizing a ANA-based electron-positron or electron-electron or $\gamma\gamma$ linear collider will only be possible with a sustained R$\&$D effort over the next two decades. Intermediate facilities are required to test the technology and demonstrate key subsystems (e.g., injector, positron generation, cooling sections, beam delivery system, final focus technology, etc.). With sufficient science motivation, one possible intermediate facility that could be considered is a 20-100 GeV center-of-mass energy advanced accelerator linear lepton collider. In what follows we focus on the science case for an intermediate facility since the necessary R\&D effort as well as potential near term applications of ANAs are outlined in other white papers submitted to Snowmass21 \cite{LPA-collider,near-term-Snowmass21,AWAKE}.

\begin{figure}
    \centering
    \includegraphics[width=0.9\columnwidth]{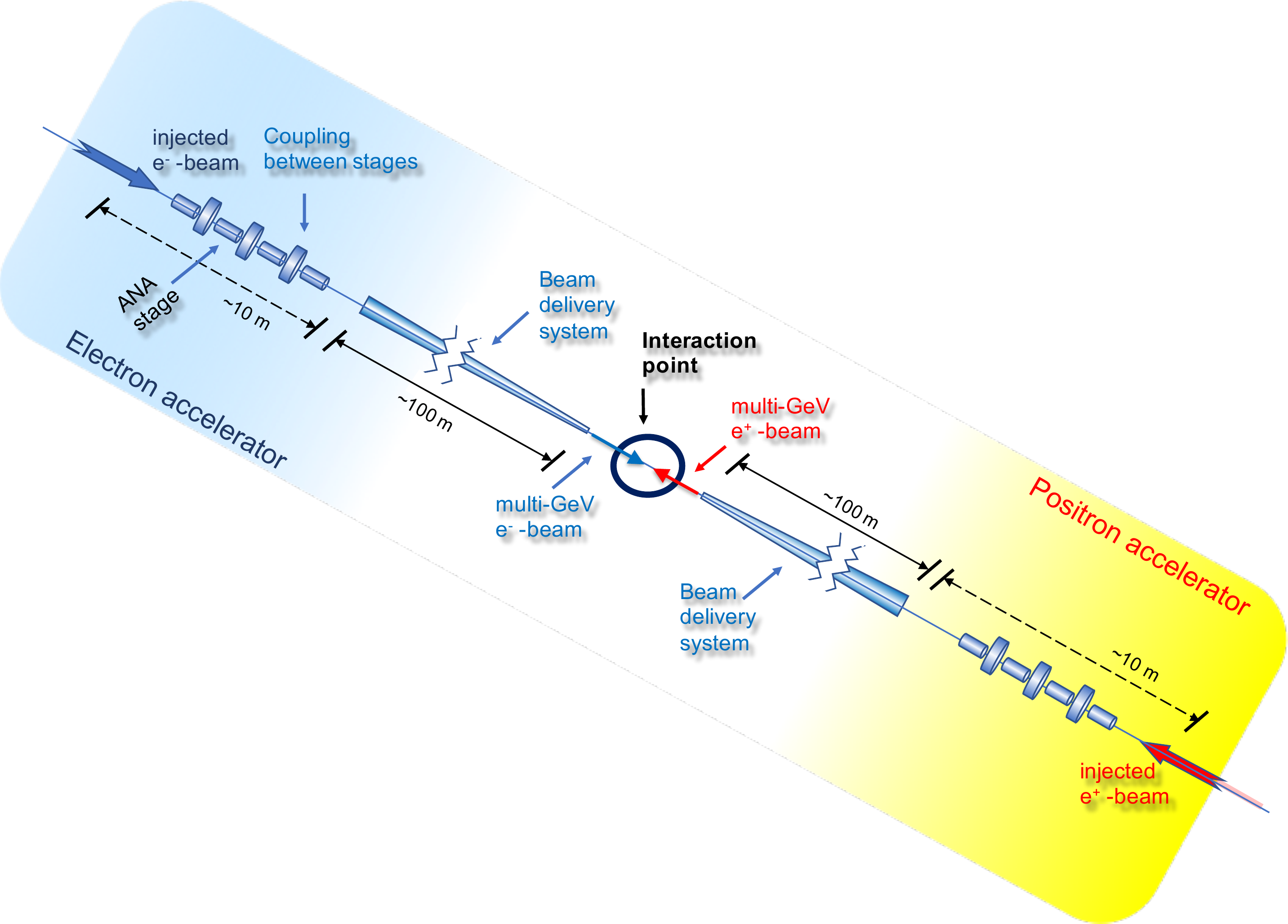}
    \includegraphics[width=0.9\columnwidth]{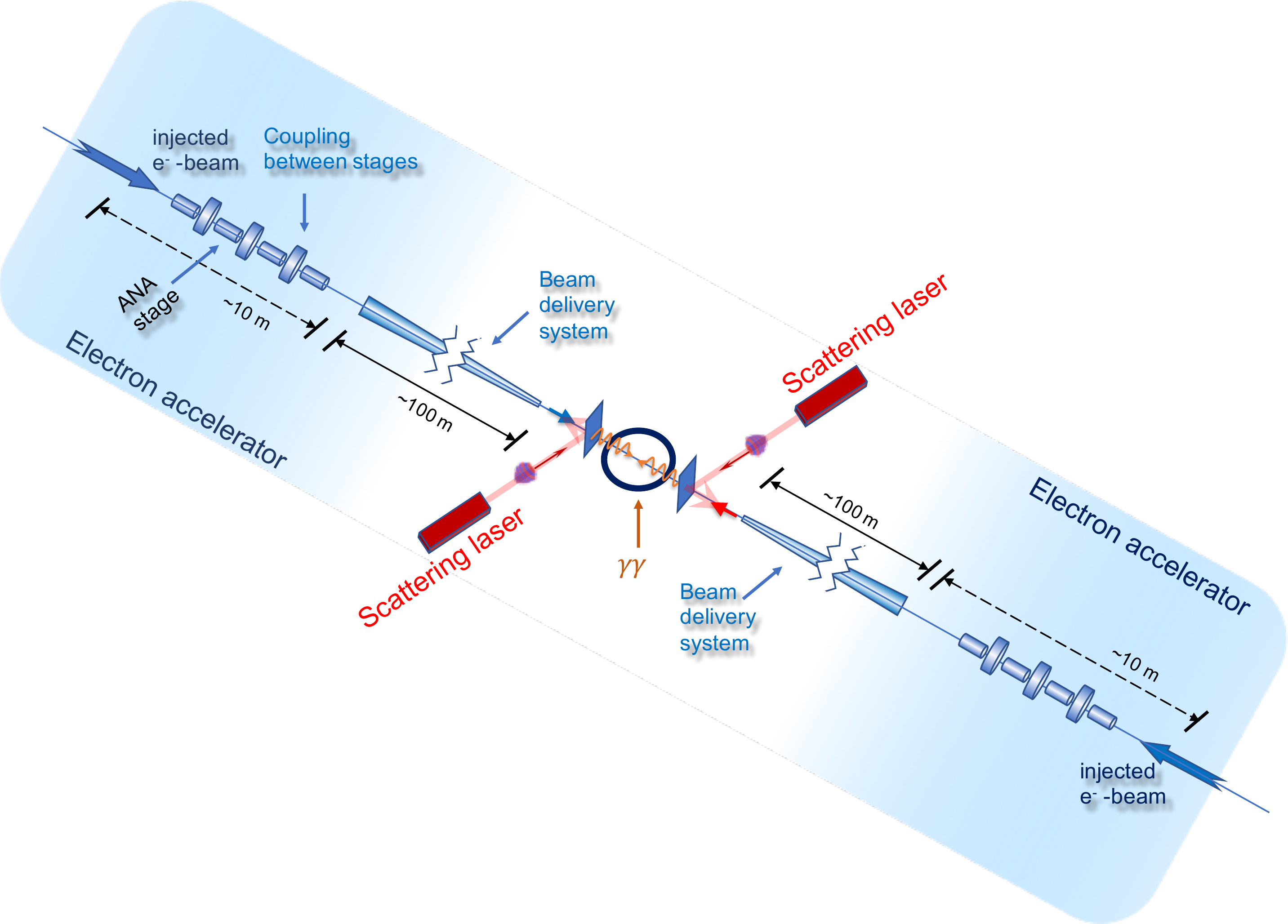}
    \includegraphics[width=0.9\columnwidth]{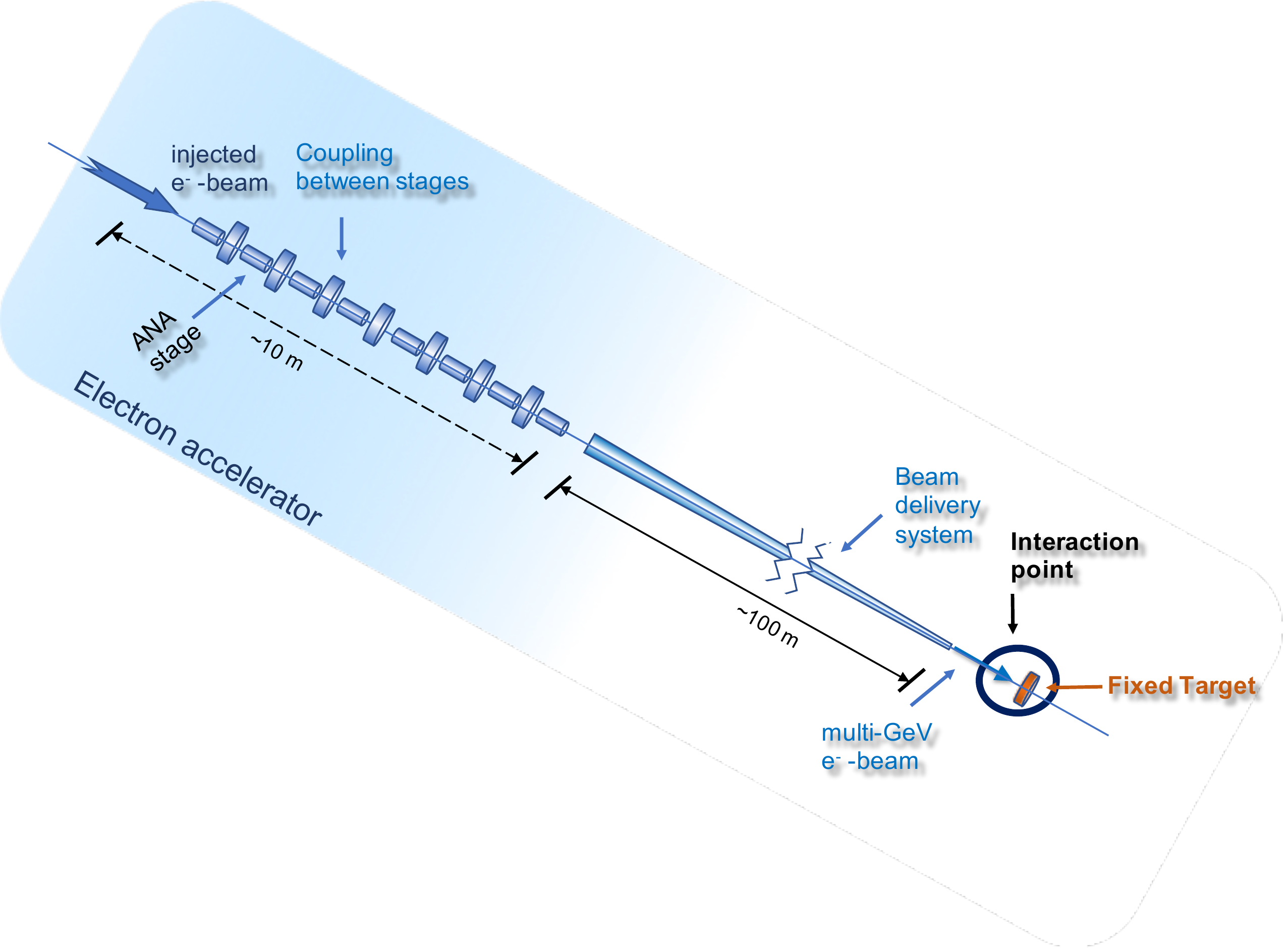}
    \caption{The principal scheme of the advanced accelerator linear collider demonstration facility at intermediate energy with possible reconfiguration into $\gamma\gamma$ collider and electron-ion collider (from top to bottom).}
    \label{fig:collider2}
\end{figure}

The science motivation for the 20-100 GeV energy range for the intermediate facility is based on the following considerations. This energy range was extensively studied more than two decades ago at different lepton colliders. However, some of the measurements can be repeated with higher statistics and using new detection techniques. The advances in theory and simulations require benchmarking against the experimental data, which might not be readily available. Below we list a number of possible directions of studies for such a facility for different center-of-mass energies: 

\begin{itemize}
\item For 0.2-1 GeV $e^+e^-$ colliders the study of $\rho$, $\omega$, and $\phi$ resonances are of most interest, however, the existing circular machines are better equipped for this type of studies. The production of true muonium, 210 MeV center-of-mass energy, which was not yet observed, might be the motivation, especially, if the required collider parameters are more attractive than those of the conventional ones.

\item For 2-5 GeV  and  7-12 GeV $e^+e^-$ colliders open avenues for the study of $\tau$ lepton and c quark, and B-meson physics respectively, which are great opportunities for QCD, exotic hadrons and search for new physics. However, the existing facilities as well as the planned upgrades for them offer parameters that far exceed the ones anticipated for the plasma based $e^+e^-$ collider. Despite this, the intermediate facility at this energy range can be used to show the possibility of muon production and acceleration. 

\item At 10-50 GeV electron beam will be the highest energy electron beam available for high energy physics studies (rare decays/dark matter searches). This is due to the fact that  nowadays the highest energy electron beam being produced is at 17 GeV at XFEL, but it is not purposed for high energy physics studies. At 45 GeV a positron beam can be used for low emittance (at the threshold) muon production on target (via $e^+e^-\rightarrow\mu^+\mu^-$). In this energy range, precision measurements related to QCD processes and searches for physics beyond the standard model  are of interest. Those include new measurements of $\alpha_s$; tests of QCD factorization and universality of hadronization effects;  benchmarking of simulation tools; searches for milli-charged particles, dark matter candidates; axion-like particles; the study of photon-photon and electron-fixed target interactions. The latter two parts of this program are synergistic with other accelerator programs, such as the Electron Ion Collider (EIC) and photon-photon collider (see Fig. 1).   

\item Further increase in energy, up to 91 GeV center-of-mass, brings the intermediate facility to the Z-pole. This energy as well as higher ones (up to 250 GeV) will, of course, be targeted by future conventional Higgs factories, which are hard to compete with. However, a cheap and compact intermediate ANA-based facility might be an attractive option given uncertain plans for the conventional facilities. 

\end{itemize}
\section{\label{sec:design} Illustrative Example: 40 GeV LPA-based collider design}

In this section we present an illustrative example of parameters of an intermediate energy demonstration facility based on LPAs. Due to the modular scheme of the facility the center-of-mass energy can be either lowered or increased to fit into the 20-100 GeV range discussed above. Moreover the nature of the facility makes it possible to re-purpose the collider for e-beam interaction with a fixed target. The role of the fixed target can also be played by a high intensity laser pulse enabling strong field quantum electrodynamics studies \cite{gonoskov.rmp.2022,zhang2020relativistic}.   

The linac of a compact, LPA-based electron-positron machine would consist of LPA stages, with each LPA stage providing 5 GeV/stage energy gain (see Fig. 2). The LPA stages, operating at a plasma density of $10^{17}$ cm$^{-3}$ accelerate 200 pC of charge with a loaded gradient of 3 GV/m. (This LPA stage would be the building block for increasing the beam energy, from tens of GeV to TeV, by adding additional stages.) To reach 20 GeV beam energy would require 4 stages in each linac arm. Taking into account the in-coupling between stages the total length of the accelerator part can be estimated to be of the order of 10 m. Each LPA stage would be powered by a laser system based on coherent combination of fiber lasers, operating at 6.5 J per pulse, 120 fs duration, and at 1 kHz repetition rate. Development of such a laser system is envisioned in the kBELLA project on the 5-10 year time scale \cite{Workshop17}.

A photocathode could be used to generate the initial polarized ($\sim$85\%) electrons, followed by bunch compressors to achieve the 8.5~$\mu$m bunch length required for coupling in the plasma accelerator. Low emittance (0.1 $\mu$m normalized) could be achieved at the electron source. Photocathodes are a mature technology that could deliver the required beam parameters. Development of laser plasma-based injectors have the potential to achieve ultra-low emittance beams beyond what is possible with conventional photocathodes, and could be explored as an electron injector option. 
The positron source would consist of the following basic scheme: a high-charge (multi-nC) beam would be generated via an LPA in the nonlinear regime; Thomson scattering of the beam would generate 10 MeV photons; The high energy photons would interact with a high-Z target to generate electron-positron pairs, which would be captured and transported to a compact damping ring; A damping ring would cool the beam to $0.1{\rm\,\mu m}$ emittance; Bunch compressors would be employed before delivering the short positron bunches to the LPA linac. At 20 GeV beam energy, a conventional beam deliver system (BDS) to the interaction point (IP) would be of the order of $\sim100\,$m, and the BDS would constitute the largest contribution to the size of the machine. 
Shorter systems based on plasma optics are also possible, but require significant R\&D. Table 1 shows possible beam parameters at the IP. Note that the LPA operates in single bunch mode, i.e., 1 ms between the arrival of the bunches, operating at 1 kHz. Assuming 30\% wall-to-laser efficiency, the overall wall-to-beam efficiency is 6\%.

\begin{figure}
    \centering
    \includegraphics[width=0.9\columnwidth]{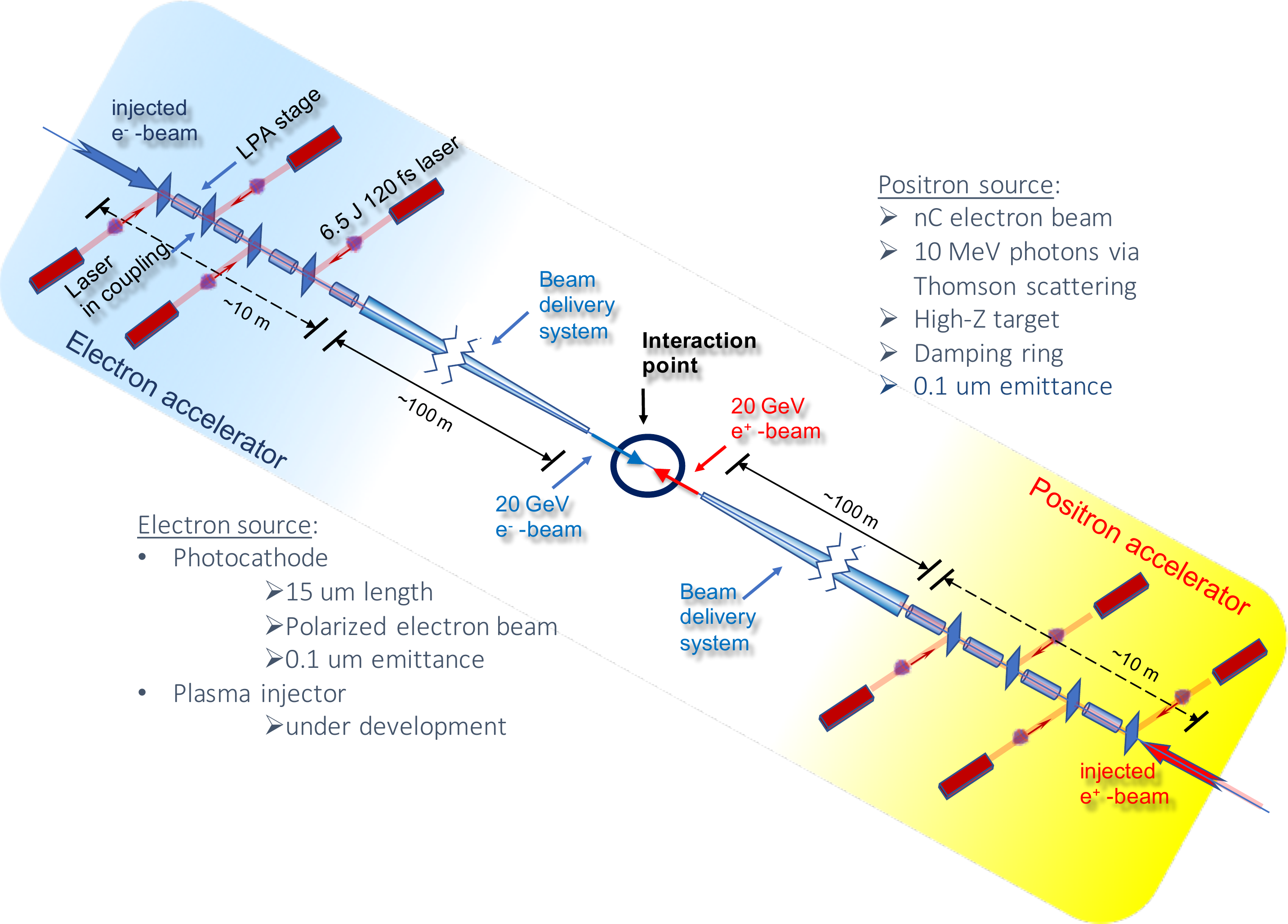}
    \caption{The principal scheme of the 40 GeV LPA-based collider.}
    \label{fig:40GeVcollider}
\end{figure}

\begin{table}
\caption{\label{tab:collider} : High-level IP parameters for $\sqrt{s} =40$ GeV $e^+e^-$ LPA-based collider.}
\begin{ruledtabular}
\begin{tabular}{lc}
Beam energy  & 20 GeV\\
Bunch charge & 200 pC\\
Bunch length (rms) & 8.5 $\mu$m \\
Repetition rate & 1 kHz (upgrade to 25 kHz)\\
Spot size at IP & 50 nm \\ 
Luminosity & $4\times 10^{30}$ cm$^{-2}$s$^{-1}$ \\ &(upgrade to $1\times 10^{32}$ cm$^{-2}$s$^{-1}$) \\
\end{tabular}
\end{ruledtabular}
\end{table}

\section{\label{sec:physics} Physics reach opportunities}

\subsection{\label{subsec:status} Multi-TeV}

First, we review the physics case for a TeV class electron-positron collider in order to put the science case for a 20-100 GeV machine in the context of the development of ANAs technology. Since the plans for the TeV class colliders are under development for the past few decades, the physics opportunities at energies up to 3 TeV are well documented. The CEPC Pre-CDR \cite{CEPC-SPPC} and the FCCee reports \cite{FCC-ee.2017,Bicer2014} detail the possible studies at at the Z resonance, the WW threshold, and the peak of the $e^+e^-\rightarrow Zh$ cross section at 240-250 GeV. The energy range of 250 GeV–1 TeV, which is relevant to the ILC, was covered in the ILC TDR \cite{ILC-TDR}, which includes both $e^+e^-$ and $\gamma\gamma$ options. Further increase in center of mass energy to 3 TeV is covered in the CLIC CDR \cite{CLIC-CDR} and in Ref. \cite{Abramowicz2017}. In what follows we list the most important goals for $e^+e^-$ and/or $\gamma\gamma$ collider in the energy range from hundred of GeV to several TeV (see \cite{ALEGRO.arxiv.2019} for details): 

\begin{enumerate}
\item The precision measurements of mass and couplings of Z and W bosons at the WW threshold \cite{FCC-ee.2017,Blondel.cern.18,baak2013}, Higgs boson coupling to the top quark and the Higgs self-coupling, top quark mass and electroweak couplings \cite{Fujii.jha.2015}. 

\item New particle searches: from precision measurements of Z and W boson masses and widths, from exotic decays of Higgs boson (possible dark matter candidates), pair-production of invisible particles (see eg., \cite{Low2014,Kowalska2018}), and new electroweak gauge bosons and lepton compositeness \cite{CLIC_Detectors}. 

\end{enumerate}

A whole new class of searches can be performed at 10 TeV–100 TeV colliders. They include the search for new interactions at short distances, which might explain the electroweak symmetry breaking, the spectrum of fermion masses, and the CP violation. For example, if we consider a 30 TeV $e^+e^-$ or $\gamma\gamma$ collider, based on advanced accelerator technology, it can offer the possibility of discovering new particles and physics beyond the SM. We can start here from the problem of the existence of the non-perturbative solution in the electroweak theory, a sphaleron, with an energy of about 9 TeV \cite{Rubakov1996,Tye.prd.2015}. 
There are a number of theoretical concepts that received a lot of attention over the years, and whose existence can be verified at a 30 TeV $e^+e^-$ or $\gamma\gamma$ collider, among them composite Higgs models \cite{weinberg.prd.1979,susskind.prd.1979}, extra space dimensions \cite{randall.prl.1999,agashe.npb.2005}, and the dynamics of flavor. In particular, if the current anomalies in the $R_K$ \cite{LHCb:2014vgu} and $R_{K^\ast}$ \cite{LHCb:2017avl} measurements at LHCb prove to be robust, a high energy collider would likely be needed to discover the particles responsible for this deviation \cite{allanach.jhep.2018,allanach.jhep.2019}. Similarly, if the muon $g-2$ anomaly \cite{Muong-2:2021ojo} withstands further theoretical and experimental scrutiny, a high energy lepton machine would be needed to robustly pin down the mechanism responsible for this effect \cite{Capdevilla:2020qel}.

\subsection{\label{sec:status} Intermediate energy demonstrator}

The proposed intermediate (20-100 GeV) lepton collider is an ideal accelerator science tool not only for studying the impact of the high current lepton beams and compact final focusing systems on the machine performance, but also for studying single particle behavior and collective plasma effects in strong electromagnetic fields. Moreover this facility can be used for proof-of-principle  acceleration of muons and protons.

The energy range of an intermediate facility was studied explicitly by PETRA (CELLO, JADE, MARK J, PLUTO, TASSO), TRISTAN (AMY, TOPAZ, VENUS), and PEP (MAC and MARK II).  Additionally, this energy was accessible via radiative return at SLC (SLD) and LEP (ALEPH, DELPHI, L3, OPAL).  The goal of this section is to explore a \textit{direct} HEP case for such a demonstrator; clearly this machine would have \textit{indirect} utility for future accelerator prototyping.

In principle, if the amount and type (e.g., polarized or not) of data for the demonstrator is the same as previous facilities and the detector is no better, then there should be no direct HEP benefit of a detector at the demonstrator.  This statement motivated two directions where a demonstrator could improve: (1) detectors could be designed to do physics that was not possible in older experiments.  For example, detected long-lived particle detectors (e.g., advanced timing/vertexing/shielding) could be uniquely sensitive to gaps in coverage by LHC and B-factories; and (2) the reality is that the data from $e^+e^-$ detectors is not generally available for reanalysis.  There have been isolated attempts to reanalyze old LEP data (e.g.,~\cite{Badea.prl.2019,chen.arxiv.2021}), but this will likely be fundamentally limited in scope due to the lack of a modern detector simulation, and access to important experimental conditions and metadata.  

Factor (2) motivates the consideration of several physics areas accessible to the intermediate collider. In particular, our understanding of QCD has significantly improved over the last two decades, with the development of new theoretical and experimental tools. Revisiting previous analyses may improve their uncertainties, and completely new measurements are now possible.  There are also searches and measurements that these collaborations did not do at the time and they could be done with a similar dataset.  In some cases, new analyses are empowered by new techniques like deep learning, that were not available at the time of LEP and earlier.  

Here are a few key examples that could be worked out in a reasonable timescale.

\begin{enumerate}

\item Precision measurements relevant to Quantum Chromodynamics: (i) strong coupling constant, $\alpha_s$ (there is an existing significant discrepancy between lattice and some $e^+e^-$ extractions, which needs to be resolved \cite{pdg.ptep.2020,Proceedings:2018jsb,Kardos2018}); (ii) Measurements to improve simulation modeling (e.g., differential measurements of fragmentation functions); (iii) New QCD measurements that were not considered previously, such as  jet substructure and groomed jet observables; (iv) and new tests of QCD factorization and universality of hadronization effects. This program will have synergies with other accelerator programs, in particular the Electron Ion Collider.

\item Laser-plasma based production and acceleration of muons. The intermediate facility is an attractive candidate for testing muon acceleration technologies. For example, at 45 GeV a positron beam can be used for low emittance (at the threshold) muon production on target (via $e^+e^-\rightarrow\mu^+\mu^-$).  

\item High energy particle interactions with strong electromagnetic fields: With the continuing increase in available laser intensity, the studies of lepton and photon interactions with electromagnetic fields are entering a new regime, which is strongly affected by strong-field quantum electrodynamics (SFQED) processes \cite{dipiazza.rmp.2012,zhang2020relativistic,gonoskov.rmp.2022,fedotov.arxiv.2022}. The hard photon emission and electron-positron ($e^+$-$e^-$) pair production significantly modify the behavior of particles giving rise to the phenomena not encountered before. For example, the coupling of these quantum processes and relativistic plasmas can result in the generation of dense $e^+$-$e^-$ pair plasma from near vacuum, complete laser energy absorption by SFQED processes or the stopping of an ultrarelativistic electron beam by an ultra-short laser pulse. The understanding of this new regime is of paramount importance for planning the next generation of ultra-high intensity laser-matter experiments and their resulting applications, such as plasma based accelerators of ions, electrons, positrons, and photons for fundamental physics studies. The SFQED experiments carried out recently, or being planned now at high power laser facilities rely on the combination of a high energy electron beam  and an intense laser pulse. However, almost every one of them will be limited in terms of achievable electron beam energy, and will have to solve the problem of stability and relatively low repetition rate. Thus, a 20-100 GeV intermediate energy demonstrator facility offers an ideal opportunity to address these issues and study the SFQED effects.    

\item Physics beyond the standard model: search for (i) the milli-charged particles; (ii) soft, displaced vertices (e.g., inelastic DM); (iii) low mass resonances decaying to hadronic final states and (iv)  axion-like particles. 

\end{enumerate}

New bosonic particles such as scalars and pseudo-scalars (axions) might be searched for by producing them through electron beam interacting with lasers, fixed targets, or a second electron beam, which speaks to the advantage of the demonstrator facility that can be reconfigured into these setups (see Fig. 1). This interaction can also be utilized to search for boson fields and milli-charged particles. When an intense electron beam collides with a laser pulse high energy photons are produced via a multi-photon Compton process \cite{dipiazza.rmp.2012, gonoskov.rmp.2022}. In the absence of new physics, the photons emitted in the electron beam-laser collision and scattered onto a physical photon dump would be detected through a decay into three photons \cite{Bai:2021dgm}. Axions would emerge in the photon dump through Primakoff effects (i.e., in scattering off nuclei) \cite{Raffelt:2006cw}. This is referred to as {\it secondary} axion production. The axion would then decay into two photons. The search for new particles would focus on the detection of correlated pairs of photons over the background of $3\gamma$ events. However, the axion can be also produced prior the interaction with the photon dump through the {\it primary} processes, such as a two step one: multi-photon Compton followed by axion emission by a photon, or a direct production of an axion in a multi-photon Compton-like processes.  Since both primary and secondary axion productions lead to the same $2\gamma$ signal, all these processes need be simulated to obtain the result. For a 20 GeV electron beam an hypothetical experiment of this type could be sensitive to the region of parameter space, which coincides with the results of similar setups including SeaQuest/DarkQuest \cite{Berlin:2018pwi, Blinov:2021say} LUXE-NPOD \cite{Bai:2021dgm}, NA62 \cite{Dobrich:2015jyk}, and FASER2~\cite{Apollinari:2015wtw}.

\subsection{\label{sec:status2} Gamma-Gamma}

It was mentioned above that the intermediate energy demonstrator facility can be reconfigured into a $\gamma\gamma$ collider \cite{telnov.loi.2021}. Though this means the addition of scattering lasers before the interaction point to generate high energy photons via Compton scattering, the $\gamma\gamma$ collider can be powered by two electron beams (see Fig. 1), which excludes the necessity to generate, cool, and accelerate positrons. The scattering laser wavelength is determined by the absence of the high-energy photon conversion into electron-positron pairs in the laser. For example, a 20 GeV electron beam needs a 0.5 $\mu$m laser wavelength to satisfy this condition. The geometric luminosities of electron-positron and $\gamma\gamma$ colliders can be similar. The development of low-emittance polarized electron sources, which is one of the goals of the ANA roadmap, can further boost the luminosity of the $\gamma\gamma$ collider.  

In the introduction we mentioned that the energy range of 2-5 GeV  and  7-12 GeV for $e^+e^-$ colliders, which is important for the study of $\tau$ lepton and c quark, and B-meson physics respectively, is well covered by the existing facilities as well as the planned upgrades for them. However, the use of $\gamma\gamma$ for these energies and above can provide a unique insight into the  spectroscopy of C = + resonances in various $J^P$ states ($b^-b$, $cc^-$, four-quark states, quark molecules and other exotic states) in a mass range exceeding that of the B-factories and not covered by any planned experiment. Moreover, variable circular and linear photon polarizations will help with the determination of quantum numbers and measurement of polarization components of the $\gamma\gamma$ cross section ($\sigma_\perp$, $\sigma_k$, $\sigma_0$, $\sigma_2$).

\subsection{\label{sec:beamdump} Beamdump}

A whole separate class of experiments arising from the electron or positron beam interaction with a special fixed target, namely, the beamdump, is usually referred to as ``beamdump experiments''. This interaction setup, given the availability of electron beams with different properties, can be sensitive to the production of axions, milli-charged particles, and light dark matter particles \cite{LOI-Beamdump}. From an accelerator physics perspective, beamdump experiments are relatively simple because it only requires a single beam for collisions and the requirements on beam quality are relaxed.

Beamdump experiments employ a variety of techniques to detect rare interactions. For the purposes of this White Paper, the most significant difference between the experiments is whether or not the search relies on kinematic measurements of beam particles. For kinematic searches, the detector attempts to reconstruct individual particle trajectories and is therefore limited to low average currents on target, but at very high repetition rate~\cite{Banerjee2020,LDMX,PADME}. Other experiments look directly for particles that are generated in and pass through the beam dump before reaching the detector~\cite{Bjorken1988,Prinz1998,Battaglieri2016}. These experiments expect high average current and are compatible with high-charge bunched beams. Some detector types work best with bunched beams because this is useful for rejecting out-of-time backgrounds~\cite{SnowdenIfft2019}. Finally, positron beams may be used to enhance production rates of dark matter particles and are compatible with a bunched-beam format~\cite{Marsicano2018}.

\begin{figure}[ht]
    \centering
        \includegraphics[width=0.95\columnwidth]{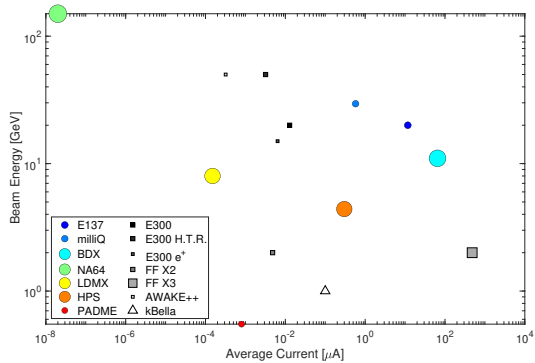}
    \caption{Colored circles: Beam energy and average current for past and planned beamdump experiments. The size of the circle is proportional to the log of the beam rate. Gray squares: Beam energy and average current of planned plasma acceleration experiments. The size of the square is proportional to the log of the beam rate.}
    \label{fig:current_v_energy}
\end{figure}

Figure~\ref{fig:current_v_energy} plots the beam energy and average current of the experiments listed in Table~\ref{tab:beam_dump} on a log-log scale. The log of the repetition rate of the experiments is represented by the size of the marker. Parameters of planned plasma wakefield acceleration experiments are shown on the same plot. Results are expected from the E300 experiment at FACET-II and the X2 experiment at FlashForward within the next two years. Results from the E300 high-transformer ratio experiment, E300 positron experiment, and X3 experiment are expected within five years. The AWAKE++ experiment will follow the currently planned AWAKE Run-II experiment which aims to demonstrate 10 GeV acceleration in the next 5 years.

\begin{table*}
\centering
\begin{tabular}{l c c c c c c c c}
    Experiment & Beam & E [GeV] & $N_b$ & Rate & $I_{avg}$ & Run Time [days] & EOT \\
    \hline
    E137 SLAC~\cite{Bjorken1988} & $e^-$ & 20 & $4\times 10^{11} $ & 180 Hz & $11.6~\mu$A & 30 & $1.8\times 10^{20} $ \\
    milliQ SLAC~\cite{Prinz1998} & $e^-$ & 29.5 & $3\times 10^{10} $ & 120 Hz & $0.576~\mu$A & 98 & $8.4\times 10^{18} $ \\
    BDX JLAB~\cite{Battaglieri2016} & $e^-$ & 11 & $1.6\times 10^{6} $ & 250 MHz & $65~\mu$A & 285 & $1\times 10^{22} $ \\
    NA64 CERN~\cite{Banerjee2020} & $e^-$ & 50-150 & $5\times 10^{6} $ & Spill & 20 fA & 90 & $1\times 10^{12} $ \\
    LDMX SLAC~\cite{LDMX} & $e^-$ & 4-8 & 20 & 46.5 MHz & 150 pA & 120 & $1\times 10^{16} $ \\
    HPS JLAB~\cite{HPS} & $e^-$ & 4.4 & $5\times 10^4$ & 40 MHz & 300 nA & 28 & $4.5\times 10^{18} $ \\
    PADME LNF~\cite{PADME} & $e^+$ & 0.55 & $1\times 10^5$ & 50 Hz & 800 fA & 250 & $1\times 10^{14} $
    
\end{tabular}
\caption{Previous and on-going beamdump-based searches. EOT refers to the total number of electrons (or positrons) on target.\label{tab:beam_dump}}
\end{table*}

\begin{figure}
    \includegraphics[width=0.9\columnwidth]{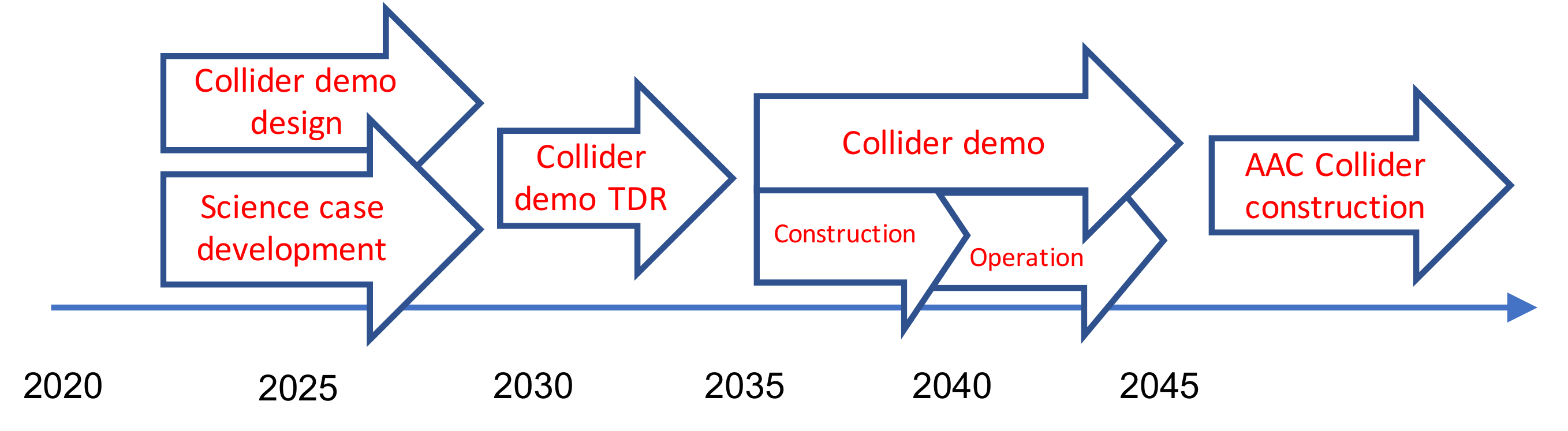}
    \caption{The timeline for the advanced accelerator linear collider demonstration facility. }
    \label{fig:collider1}
\end{figure}

\section{Summary}

Realizing the challenge of a TeV collider based on advanced and novel accelerating techniques will only be possible with a sustained, decades-long R\&D effort. Intermediate facilities will be required to demonstrate key accelerator technologies and subsystems that are compatible with the ANA technology. One may consider a 20-100 GeV collider design, as a possible intermediate facility. With sufficient science motivation, such a machine could be pursued, enabling key components to be tested on the path to a TeV collider, as well as training
of the next generation of accelerator physicists and engineers.

In view of the above considerations regarding possible science case for such a facility the advanced accelerator and HEP communities propose the following recommendations to the Snowmass conveners:

\begin{enumerate}
\item The research continue on the science case and for the intermediate facility in the framework of the General Accelerator R\&D program.

\item A design to be carried out for a  collider demonstration facility at an intermediate energy (20-100~GeV)  to test the technology and demonstrate key subsystem, as well as  provide a facility for physics experiments at intermediate energy.   

\end{enumerate}

These recommendations are parts of the timeline (see Fig. 4 and \cite{LPA-collider} for a TeV collider timeline), which envisions the development of the science case and collider demo design by 2030, which will feed into the technical design report. 

\acknowledgments
This work was supported by the Director, Office of Science, Office of High Energy Physics, of the U.S. Department of Energy under Contract No.\ DE-AC02-05CH11231. The work of CL was supported by the U.S. Department of Energy, Office of Science, through the Office of Nuclear Physics and with an Early Career Research Award at Los Alamos National Laboratory, which is operated by Triad National Security, LLC, for the National Nuclear Security Administration of the U.S. Department of Energy under Contract No.~89233218CNA000001.

\end{document}